# PHOTON-DETECTOR ENTANGLEMENT BY INTERFERENCE AT BEAM SPLITTER (COLLAPSE AS SPONTANEOUS SUPERPOSITION BREAKING)

Vladan Panković

Department of Physics, Faculty of Sciences, 21000 Novi Sad,
Trg Dositeja Obradovića 4, Serbia, vladan.pankovic@df.uns.ac.rs

## Abstract

In this work we suggest an original, realistic and simple, form of the experiment of interference of single photon at beam splitter. Here quantum entanglement (Schrödinger cat effect) between photon and photon trajectories detector (which includes two same simple two-level systems) is studied. It admits a consistent model of the detection or measurement procedure. Concretely here (self-)collapse at detector is consistently treated as the spontaneous (non-dynamical) breaking (effective hiding) of the unitary symmetry (superposition or entanglement) of quantum mechanical dynamics. It can be considered as an especial case of the general formalism of the spontaneous symmetry breaking.

## 1. Introduction

In this work we shall suggest an original, realistic and simple, form of the well-known experiment of interference of single photon at beam splitter. Here quantum entanglement (Schrödinger cat effect) between photon and photon trajectories detector (which includes two same simple two-level systems, practically two atoms) will be studied. It admits a consistent model of the detection or measurement procedure, i.e. (self-)collapse. Concretely here (self-)collapse at detector is consistently treated as the spontaneous (non-dynamical) breaking (effective hiding) of the unitary symmetry (that conserves superposition or entanglement) of quantum mechanical dynamics. It can be considered as an especial case of the general formalism of the spontaneous symmetry breaking.

As it is well-known [1]-[6] experiment of the interference of single quantum system (photon, electron, etc.) at a diaphragm with two slits or, analogously, at beam splitter represents corner stone for demonstration of the basic principles of the standard quantum mechanical formalism [2], [3], [7]-[9]. In this experiment, simply speaking, source S emits single quantum system (with well-defined momentum or corresponding de Broglie wave-length) that propagates toward beam splitter BS. In time moment chosen to be initial quantum system exactly quantum dynamically deterministically interacts with beam splitter. As the exact result of this interaction quantum state of the quantum system evolves exactly deterministically during time in the following way. In time moment t after this interaction quantum system is described by quantum superposition of quantum reflected state, i.e. reflected trajectory $|R(t)\rangle$ and quantum transmitted state, i.e. transmitted trajectory $|T(t)\rangle$

(1) $$i\, 2^{-1/2} |R(t)\rangle + 2^{-1/2} |T(t)\rangle$$

Reflected and transmitted state will be later redirected by two total mirrors MR and MT (at the same distance in respect to beam splitter) respectively toward the same detection photo plate DPP. For this reason it will be considered that total reflected trajectory is determined, roughly speaking, by centers of BS,MR and DPP, while total transmitted trajectory is determined by centers of BS,MT and DPP. In any individual case detection photo plate detects finally always only one photon. But at a statistical ensemble of the individual detections statistical distribution of the photons along detection photo plate axis can be different.

If between beam splitter and detection photo plate quantum system does not interact with any other physical system able to detect concrete trajectory of quantum system then detection photo plate finally detects statistical distribution with characteristic interference terms (existence of many local minimums and maximums of the distribution function) corresponding to superposition (1) as the exact result of the exact quantum dynamical evolution, or, roughly speaking, to propagation of the quantum system similarly as a wave (Figure 1.). It is very important to be pointed out that here detection photo plate realizes detection at the quantum system as an isolated system.

But if between beam splitter and photo plate quantum system interacts with such physical system, called formally collision system, detector or pointer, D, able to detect concrete trajectory of quantum system then photo plate finally detects statistical distribution with effective absence of the interference terms (distribution function with one maximum and without any minimum): It corresponding to principally probabilistic breaking (collapse) of superposition (1) in statistical mixture of reflected and transmitted trajectory with statistical weights $|i\ 2^{-1/2}|^2$ and $|2^{-1/2}|^2$ respectively (Figure 2.). Of course such collapse cannot be exactly presented as the exact result of the exact, deterministic quantum dynamical evolution. Generally speaking, according to standard quantum mechanical formalism, exact quantum dynamical evolution must be unitary symmetric. It simply means that it conserves superposition so that quantum dynamical state of the quantum system can be with the same right presented as corresponding superposition in any basis of the Hilbert space of the quantum states. It means, since a basis in Hilbert space represents corresponding referential frame ("observer") in Hilbert space, that all referential frames in Hilbert space have the same rights in description of the exact quantum mechanical dynamics. (Really, we can generally with the same right, use Heisenberg matrix picture, Schrödinger wave picture, interaction picture, Ehrenfest average value picture, Dirac picture, Feynman path integral picture, etc. even if in some cases one of mentioned pictures can be technically more appropriate for calculations than other.) Absence of the absolute referential frame (absolute "observer") in exact quantum dynamics and its conceptual analogy with the absence of the absolute referential frame in theory of relativity has been many times pointed out by Bohr [1] [10].

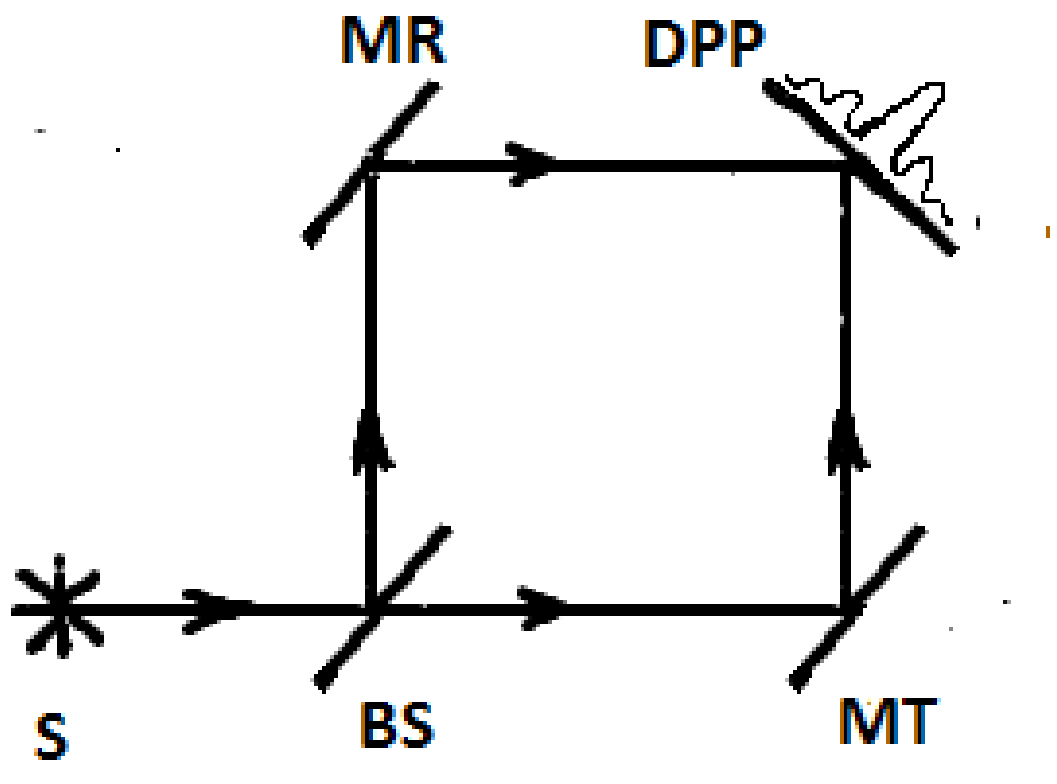

Figure 1.

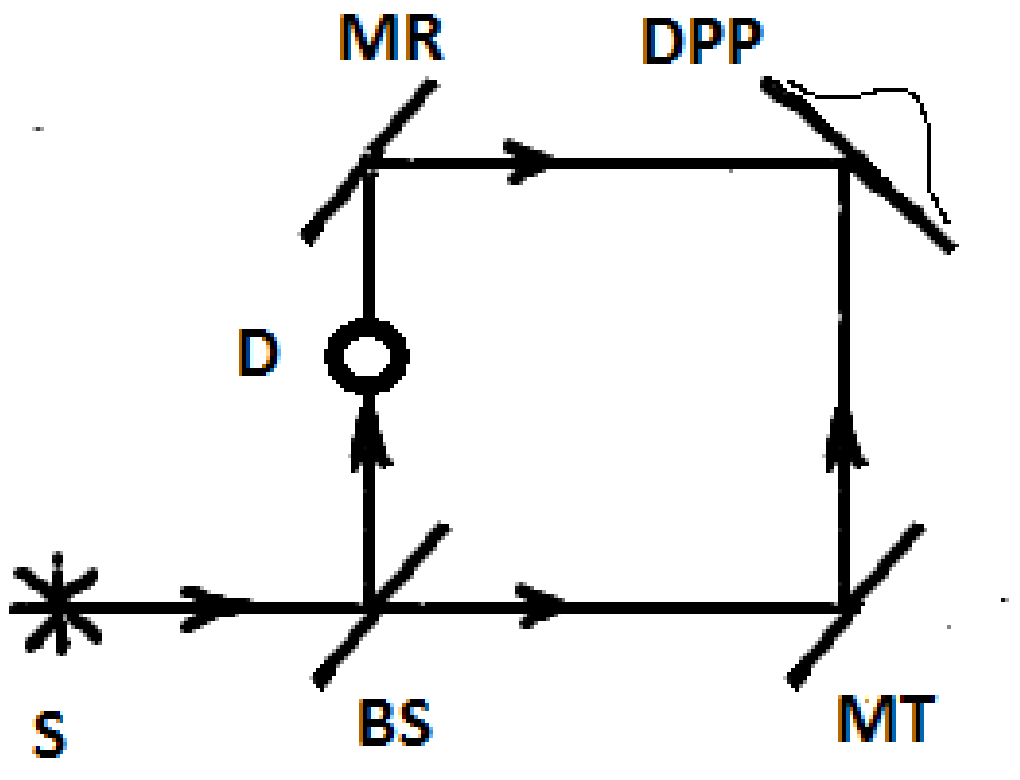

Figure 2.

All this, however, does not represent necessarily a contradiction within standard quantum mechanical formalism. Strictly quantum dynamically speaking interaction between quantum system and collision system realizes deterministically an entangled quantum state of the quantum super-system which includes both, quantum system and collision system. Simply speaking, by mentioned quantum dynamical interaction superposition (1) of the quantum system is extended in entangled quantum state of the super-system, i.e. in the super-systemic superposition. Also, as well as superposition (1) does not admit absolute separation of the reflected and transmitted state, entangled quantum state, as it is well-known [10], [11] , does not admit absolute separation of quantum super-system in quantum system and collision system described by quantum states or statistical mixtures of quantum states. Precisely, standard quantum mechanical formalism [8] admits that the same entangled state of quantum super-system be formally separated in different (complementary) ways (different bi-orthogonal sub-systemic expansions) by correlated so-called second kind statistical mixtures of the quantum system and collision system. None of mentioned different formal separations is predominant or absolute within entangled state of quantum super-system. For this reason Everett "relative state" quantum mechanics [12] is essentially incomplete within standard quantum mechanical formalism. On the other hand only a new detector can relatively chose some of mentioned different formal separation most appropriate for new detection, but it, within standard quantum mechanical formalism cannot be considered as any absolute delayed choice or change of the past (supposed by Wheeler [5]). In this way existence of the entangled quantum states within standard quantum mechanical formalism means that strictly deterministic, unitary symmetric quantum dynamical evolution can be consistently applied for description of the quantum systems as well as quantum super-systems. It is in an excellent agreement with experimental data [13] , [15] that point out (oppositely to different decoherence theories [16]) that quantum entanglement exists (in absence of additional interactions with additional quantum systems) independently of the distance or temperature. It is very important to be pointed out that mentioned independence within standard quantum mechanical formalism does not imply any "super-luminal spooky action at distance" between quantum systems. Exact quantum mechanical dynamics is relativistically local. Bell inequality [17] whose breaking in real experiments [13] , [14] implies super-luminal interactions does not refer on the unitary symmetric quantum mechanical dynamics but on the such hypothetical non-unitary symmetric generalized dynamics, or such generalized dynamics within which unitary symmetry is dynamically broken, within different so-called hidden variables theories [18]- [20]. For this reason such theories must be either inconsistent with experimental data or super-luminal and in this sense implausible. Finally, after quantum dynamical interaction between quantum system and collision system, i.e. after realization of the entangled quantum state at quantum super-system, detection plate interacts with quantum system as sub-system of quantum super-system. For this reason here detection results, i.e. collapse can be effective but not absolute. In other words, according to standard quantum mechanical formalism [8], before detection that detection photo plate should to realize at quantum system, quantum system is described either by first kind statistical mixture of the trajectories (if absolute collapse has been realized by first detection by collision system) or by formal second kind mixture of the trajectories (if by interaction between quantum system and collision system entanglement is realized). Since both mentioned, first and second kind statistical mixture, have identical mathematical form, mentioned second detection that will be done by detection photo plate cannot decide what really appears, absolute collapse or entanglement. Any supposition that absolute collapse appears (including von Neumann hypothesis that absolute collapse is generated by Abstract Ego of the human observer [9]) leads necessarily toward some kind of hidden variables theories (either inconsistent or implausible). Opposite supposition, that exact quantum dynamical interaction between quantum system any collision system realizes entanglement, needs additional formalization of the collapse by measurement or detection. First step in this direction are ideas [21], [22] that collapse in some sense represents an approximate phenomena connected with weak interference conditions. However, as it is well-known, at the exact quantum mechanical level of analysis accuracy, superposition of weakly interfering quantum states stands superposition, without any approximate transition in statistical mixture. Second, very important step in mentioned direction was observation (without detailed physical explanation) [23] that mathematical structure of the standard quantum mechanical formalism admits that collapse be considered as a continuous Landau phase transition. It can be correlated with third step in the same direction, i.e. with old Bohr ideas [1], [10] on approximate “classical” description of the measurement, i.e. detection apparatus in detection procedure, when an approximate level of analysis accuracy must be introduced. All this admits, as it has been definitely proved [24], [25], that collapse can be considered as the spontaneous (non-dynamical) breaking (effective hiding) of the unitary symmetry (superposition or entanglement) by phase transition from exact quantum in approximate level of the analysis accuracy. It can be considered as an especial case of the general formalism of the spontaneous symmetry breaking that can be successfully applied in many different domains of the physics, e.g. elasticity of rigid bodies, quantum theory of ferromagnetism, and, especially, in quantum theory of electro-weak interactions [26]-[28].

Model of the collapse as spontaneous superposition breaking we shall now apply in discussion of the interference of single quantum system at beam splitter. But more detailed analysis of this question needs firstly a concrete model of the collision system.

## 2. A photon-detector entanglement existence experiment

It is not necessary that collision system be a macroscopic system. Namely, it is well-known [4] that by interference of single electron at diaphragm with two slits collision system can be in principle even single photon, but of course technically is much simpler that experiment be realized with many photons, i.e. macroscopic ensemble of the photons.

Suppose that quantum system represents a photon P while collision system represents two distant, identical, simple two-level systems (practically two atoms) TLS1 and TLS2 with excited state |E> and ground state |G> (Figure 3.). Suppose that difference between energy of the excited state and energy of the ground state is equal to photon energy. Also suppose that life time of the excited state of two-level system is much larger than time propagation of the photon between source, beam splitter and detection photo plate. Finally suppose that motion of any two-level system during photons propagation can be neglected effectively. Such collision system with TLS1 in excited state and TLS2 in ground state will be initially placed, roughly speaking, at photon reflected trajectory in the middle between beam splitter and MR

If photon propagates along reflected trajectory its exact quantum dynamical interaction with first two-level system TLS1in excited state can be considered as the stimulated emission of a new, copy photon coherent with initial, without any change of the reflected state of the initial photon. Simultaneously TLS1 turns out in the ground state. Later two photons that propagate coherently along reflected trajectory exactly quantum dynamically interact with second two-level system, It can be considered as the absorption of one photon by TLS2 which turns out in excited state without any change of the reflected state of other photon. In this way, after interaction with photon(s), both two-level systems change their initial states. Finally, after redirection by total mirror MR, photon propagates toward detection photo plate DPP where it will be detected.

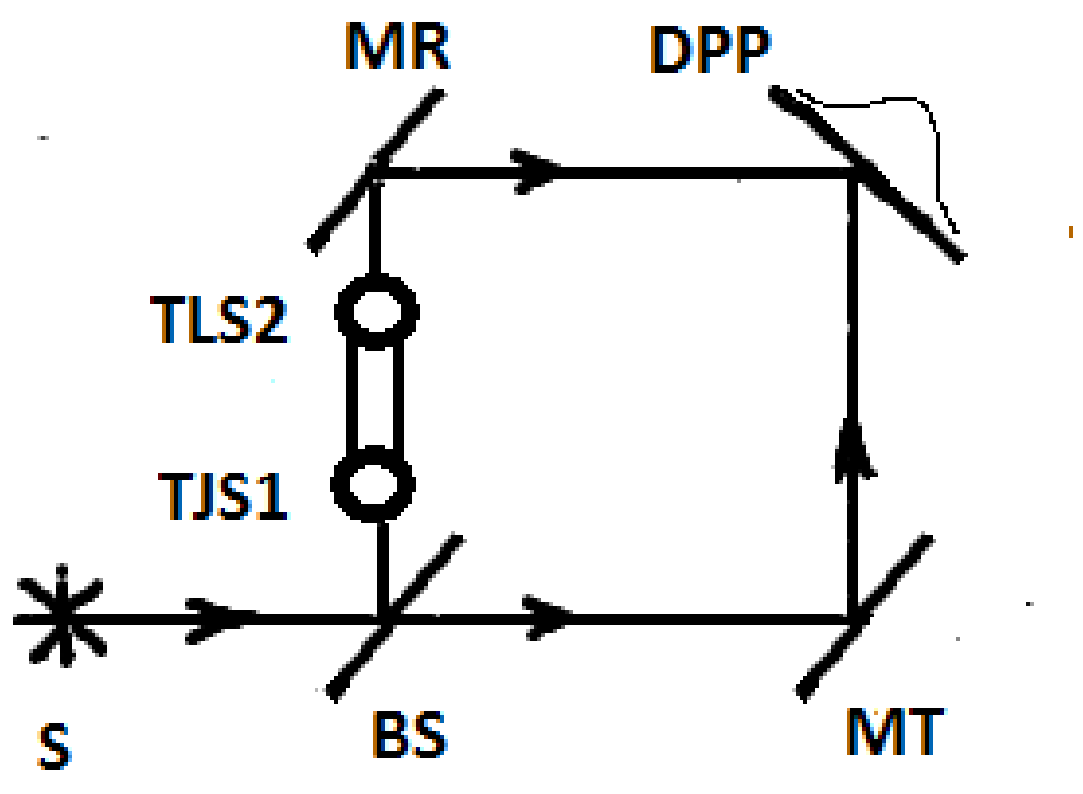

Figure 3.

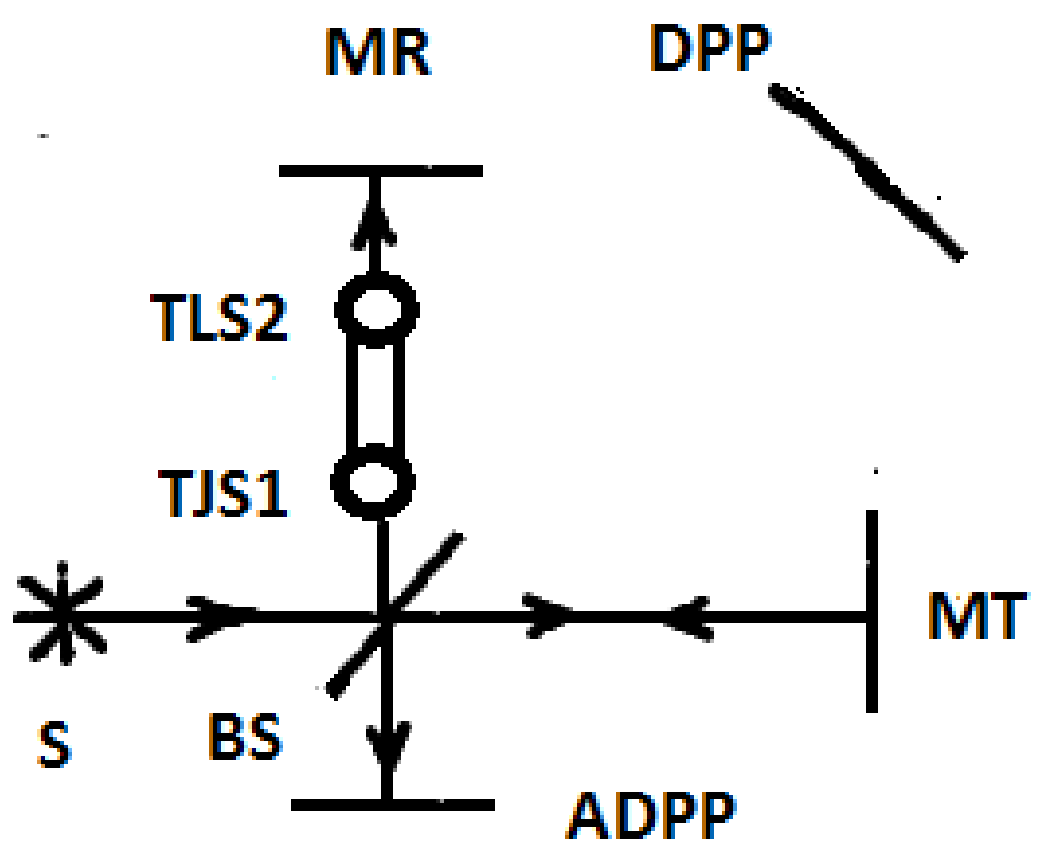

Figure 4.

If photon propagates along transmitted trajectory it exactly quantum dynamical does not interact with any of two two-level system that stands in the initial states. Finally, after redirection by total mirror MT, photon propagates toward detection photo plate DPP where it will be detected.

In this way it is definitely proved that mentioned two two-level systems can be consistently treated as collision system, detector or pointer of the photon trajectory. Also it can be repeated and pointed out the following. By mentioned detection neither reflected nor transmitted trajectory of the photon will be changed.

So, according to suppositions, before any quantum dynamical interaction between P, TLS1 and TLS2, for P jn superposition state (1), TLS1 in excited and TLS2 in gr4ound state, quantum super-system P+TLS1+TLS2 (that includes all P, TLS1 and TLS2) is described by the following non-entangled quantum state

(2) $$( i\, 2^{-1/2} |R(t)\rangle + 2^{-1/2} |T(t)\rangle) |E\rangle|G\rangle$$

As it is well-known, in distinction to totally stable ground state excited state is dynamically non-stable. Quantum system in excited state spontaneously decays during time, i.e. it can spontaneously emit single photon and turn out in the ground state. It means that excited state during its initial stability represents in fact bound state of the quantum system and photon localized within this quantum system. (This fact can be approximately neglected in usual quantum mechanical considerations [7] but in some cases, especially by transition from quantum mechanics in quantum field theory, it can have principal role which will be discussed later.) For this reason, according to introduced suppositions, expression (2) can be presented in the following form

$$(3)\qquad ( i\ 2^{-1/2}\ |R(t)> + 2^{-1/2}\ |T(t)>)\ |P,TLS1>\ |\ G\ >|G>$$

where |P,TLS1> represents quantum state of single photon localized within TLS1.

Immediately after exact quantum dynamical interaction between P and TLS1 non-entangled quantum state (2) or (3) deterministically evolves in the following, partially entangled state

$$\begin{aligned}(4)\qquad & ( i\ 2^{-1/2}\ |R(t)>\ |R(t)>\ |G> + 2^{-1/2}\ |T(t)>\ |E>)\ |G> = \\ & ( i\ 2^{-1/2}\ |R(t)>\ |R(t)>\ |G> + 2^{-1/2}\ |T(t)>\ |P,TLS1>\ |G>)|G>\end{aligned}$$

It, simply speaking, describes entanglement of two photons so that first entanglement term describes both photons that free propagate along reflected trajectory while second entanglement term describes one photon that free propagates along transmitted trajectory and other photon localized within TLS1.

Immediately after exact quantum dynamical interaction between P+TLS1 and TLS2 partially entangled quantum state (4) quantum dynamically deterministically evolves in the following (partially) entangled state

$$\begin{aligned}(5)\qquad & i\ 2^{-1/2}\ |R(t)>\ |G>\ |E> + 2^{-1/2}\ |T(t)>\ |E>|G> = \\ & ( i\ 2^{-1/2}\ |R(t)>\ |P,TLS2>\ |G>\ |G> + 2^{-1/2}\ |T(t)>\ |P,TLS1>|G>\ |G>) = \\ & ( i\ 2^{-1/2}\ |R(t)>\ |P,TLS2> + 2^{-1/2}\ |T(t)>\ |P,TLS1>)\ |G>\ |G>\end{aligned}$$

It, simply speaking, describes entanglement between free propagating photon and both two-level systems. But it can be presented in form of the partially entangled state within which free propagated and localized photon are entangled while both two-level systems are non-entangled.

Now we can discuss obtained result, i.e. expression (5). First of all, according to standard quantum mechanical formalism, detection photo plate must realize detection at entangled quantum state (5), or formally, at entangled state of two photons

$$(6)\qquad i\ 2^{-1/2}\ |R(t)>\ |P,TLS2> + 2^{-1/2}\ |T(t)>\ |P,TLS1>$$

by sub-systemic detection at free propagating photon without any immediate detection of the localized photon. Free propagating photon holds two quantum states or two trajectories, reflected |R(t)> and transmitted |T(t)>, which between beam splitter and detection photo plate weakly interfere, but which nearly detection photo plate strongly interfere. Localized photon holds two quantum states |P,TLS1> localized nearly TLS1 and |P,TLS2> localized nearly TLS2, which, for sufficiently distant TLS1 and TLS2, extremely weakly interfere. Finally, it can be observed that according to standard quantum mechanical formalism [8], second kind statistical mixture for free propagating photon, for entangled state (6) and supposed form of the new photo plate detection, holds statistical mixture of reflected and transmitted trajectory with statistical weights $|i\ 2^{-1/2}|^2$ and $|2^{-1/2}|^2$ respectively

Existence of the entangled state (6) can be testified experimentally very simply as it is presented at Figure 4. . Here both total mirrors MR and MT are rotated in such way that now redirected reflected and transmitted photon go back toward beam splitter. Such rotations of the total mirrors do not any influence at the realization of the entanglement (6) (if it is really realized) by previous quantum dynamical interactions between photon and two two-level system.

Immediately after redirection of the photon by total mirrors entangled state (5) quantum dynamically deterministically evolves in the following entangled state

(7) $i\, 2^{-1/2} (i|AR(t)\rangle) |G\rangle |E\rangle + 2^{-1/2} (i|AT(t)\rangle) |E\rangle|G\rangle$

Here $|AR(t)\rangle$ describes propagation of the photon along reflected trajectory in opposite direction while $|AT(t)\rangle$ describes propagation of the photon along transmitted trajectory in opposite direction. Since standard quantum mechanical formalism admits neglecting of the i phase shift expression (7) can be simplified and changed by

(8) $i\, 2^{-1/2} |AR(t)\rangle |G\rangle |E\rangle + 2^{-1/2} |AT(t)\rangle |E\rangle|G\rangle$

Further, immediately after interaction of anti-reflected photon and TLS2 in ground state entangled state (8) quantum dynamically deterministically evolves in the following partially entangled state

(9) $i\, 2^{-1/2} |AR(t)\rangle |AR(t)\rangle |G\rangle |G\rangle + 2^{-1/2} |AT(t)\rangle |E\rangle|G\rangle =$
$(i\, 2^{-1/2} |AR(t)\rangle |AR(t)\rangle |G\rangle + 2^{-1/2} |AT(t)\rangle |E\rangle) |G\rangle$

Obviously, here TLS2 in ground state is non-entangled with other systems. Entangled state of mentioned other systems $(i\, 2^{-1/2} |AR(t)\rangle |AR(t)\rangle |G\rangle + 2^{-1/2} |AT(t)\rangle |E\rangle)$ holds two terms. First term corresponds to propagation of two coherent anti-reflected photons toward TLS1 in ground state. Second term corresponds to propagation of anti-transmitted photon toward beam splitter.

Further, immediately after interaction of two anti-reflected photons and TLS1 in ground state partially entangled state (9) quantum dynamically deterministically evolves in the following non-entangled state

(10) $(i\, 2^{-1/2} |AR(t)\rangle + 2^{-1/2} |AT(t)\rangle) |E\rangle|G\rangle$

Here TLS1 is in excited state, TLS2 is in ground state, while single photon propagates toward beam splitter in the following superposition state

(11) $i\, 2^{-1/2} |AR(t)\rangle + 2^{-1/2} |AT(t)\rangle$

Immediately after interaction between photon and beam splitter superposition (11) quantum dynamically deterministically evolves in the following quantum state

(12) $i\, 2^{-1/2} ( i\, 2^{-1/2} |\leftarrow (t)\rangle + 2^{-1/2} |\downarrow (t)\rangle ) + 2^{-1/2} (2^{-1/2} |\leftarrow (t)\rangle + i\, 2^{-1/2} |\downarrow (t)\rangle ) =$
$= i|\downarrow (t)\rangle$

where $|\leftarrow (t)\rangle$ describes photon that propagates toward source while $|\downarrow (t)\rangle$ describes photon that propagates toward additional detection photo plate ADPP (under beam splitter). It implies that photon certainly, i.e. with probability $|i|^2 = 1$, propagates toward additional detection plate if entangled state (5) really exists.

In case that instead of entangled state (5) corresponding statistical mixture, as consequence of the absolute collapse, really exists then form of the statistical mixture of quantum states will be conserved in all further interactions between photon, two-level systems and beam splitter, so that, as it is not hard to see, photon will finally propagate toward detection plate with probability 1/2 only.

Until realization of mentioned experiment of the existence of the photon-two-level systems entanglement it can be expected (we paraphrase Einstein) as the best chance for the Nature, the following. Since entanglement existence are really experimentally verified in a wide class of the quantum phenomena under various conditions and since in all steps of mentioned theoretical considerations on the photon-detector entanglement only simple, well-known quantum dynamical effects are consistently used, it can be expected that photon-detector entanglement will be unambiguously experimentally verified. In this case additional formalization of the collapse by measurement or detection is necessary. For this reason we shall consider here model of the (self-)collapse at detector as the spontaneous (non-dynamical) breaking (effective hiding) of the unitary symmetry (superposition or entanglement) of quantum mechanical dynamics.

## 3. Basic concepts of the general formalism of the spontaneous (non-dynamical) symmetry breaking (effective hiding)

As it is well-known [26]-[28] there are two principally different ways for breaking of the dynamical symmetry, dynamical and spontaneous (non-dynamical), which will be shortly considered.

Basic characteristic of a physical theory that can be applied for description of a physical system is corresponding dynamics. This dynamics or precisely dynamical equations hold corresponding dynamical symmetries. Unique solution of mentioned dynamical equations, if it exists, represents dynamically stable and observable dynamical state that deterministically evolves during time. (Words stable and observable here refer to corresponding dynamics.)

Consider situation in which the physical system can be described by two discretely different physical theories, one, more accurate or simply speaking exact, and other, less accurate or simply speaking approximate. (In this sense we can speak about two different levels of analysis accuracy, one exact corresponding to exact theory and other approximate corresponding to approximate theory.) Exact theory can be considered as a non-trivial extension of the approximate theory, while approximate theory can be considered as a non-trivial reduction of the exact theory.

There is such situation in which some approximate dynamical symmetry does not represent any exact dynamical symmetry. It means that mentioned approximate dynamical symmetry is not conserved at exact level of analysis accuracy or that mentioned approximate dynamical symmetry becomes broken by transition from approximate at exact level of analysis accuracy. It represents dynamical breaking of the approximate dynamical symmetry. Dynamical symmetry breaking holds many very important applications within physics, classical and quantum. For example, experimentally verified parity breaking in weak interactions is a typical dynamical symmetry breaking. But there are such very important physical situations where concept of the dynamical symmetry breaking cannot be applied at all. For example in attempt of unification of the electromagnetic and weak interaction additional mass term cannot be immediately introduced in the dynamics since such introduction break dynamical gauge symmetry without which theory does not admit renormalization and diverges

There is such situation in which exact dynamics should be approximately reduced or projected in approximate dynamics. For reason of the discrete difference between exact dynamics and approximate dynamics there are different possibilities for realization of mentioned projection. Some exact dynamical state can be globally (i.e. in whole space of the arguments) consistently (convergently) approximated by corresponding approximate dynamical state and then some exact dynamical symmetries become effectively hidden but not broken. Here, roughly speaking, exact dynamical state becomes globally dynamically stable and observable even at the approximate level of analysis accuracy. Some other exact dynamical state that cannot be globally (i.e. in whole space of the arguments) approximated by any corresponding approximate dynamical state. For this reason, exactly existing, exactly dynamically stable and observable dynamical state is globally approximately dynamically non-stable and non-observable. Here, roughly speaking, exact dynamical state becomes globally non-stable and non-observable at the approximate level of analysis accuracy. Finally, there is such exact dynamical state cannot be globally (i.e. in whole space of the arguments) but can be locally (in some disjunctive domains of the arguments) consistently (convergently) approximated by corresponding local approximate dynamical states. In this sense such exact dynamical state is locally approximately dynamically stable and observable. It means that within any mentioned local domain of arguments approximate dynamics holds one consistent solution, i.e. approximate dynamical state which within this domain can locally represent exact dynamical state. Formally we can speak about spontaneous (non-dynamical) transition from the globally approximately dynamically non-stable exact dynamical state in the locally approximately dynamically stable exact dynamical state. But this transition or this event is inherently probabilistic or statistical. Namely it cannot be described deterministically neither by exact dynamics (within which such transition does not exists at all), nor by approximate dynamics (since its validity is limited in restricted domain of the exact dynamical state arguments). According to usual geometric definition of probability, probability of mentioned transition or event can be defined as the relative measure of corresponding arguments domain. It can be added that here exact dynamical state cannot be simultaneously separated in all local domains in case when its norm must be conserved as it is case within quantum field theory. Finally, after mentioned probabilistic local transition or projection, further deterministic approximate dynamical evolution appears which does not admit reverse transition. Discussed transition by which many, symmetric, local approximate dynamical solutions are probabilistically and ireversiblely changed by one actual local approximate dynamical solution represents in fact spontaneous symmetry breaking. More accurately speaking we have here effective hiding of the symmetry at approximate level of analysis accuracy (at exact level of analysis accuracy symmetry is not broken but it is conserved).

Spontaneous symmetry breaking holds many very important applications within physics, classical and quantum. Within electro-weak theory exact quantum dynamical solution of the exact dynamical equation exists, but it cannot be obtained analytically. For this reason approximate dynamics, i.e. perturbation theory must be used. Perturbation theory diverges for zero field value (false vacuum), but locally converges for “circularly” distributed field values (real vacuums). Translation, i.e. transition from zero field value (false vacuum) to some field value at “circle” (real vacuum) realizes spontaneous breaking of the “circular symmetry” (in fact gauge symmetry). As it has been detailedly discussed in [28], word “translation” here is not conclusive, as well as word “choice” or question “how Nature chose one of equally probable real vacuums”. All mentioned phrases refer on the dynamical breaking of the symmetry concept. Within spontaneous symmetry breaking there is no “choice” but only irreducibly probabilistic event at approximate, perturbation theory level of accuracy, and exactly conserved gauge symmetry. In domains of not so high energies theory of perturbation can be used as technically simple theoretical method, but it cannot be considered as any principal “choice”.

## 4. Collapse as spontaneous (non-dynamical) superposition breaking (effective hiding)

As it is not hard to see according to previous discussions all theories of the measurement or detection which suppose existence of absolute collapse or hidden variables are based on the concept of the dynamical breaking of the unitary symmetry, i.e. dynamical breaking of the superposition or entanglement. According to experimental data all such theories are either inconsistent (if such theories are local) or implausible (if such theories are super-luminal or non-local). Implausibility of such super-luminal theories follows from fact that such theories do no admit existence of classical mechanics as a limit of quantum mechanics (but suggest a non-trivial extension of both quantum and classical mechanics within some super theory), existence of the classical field theories that are local, and, existence of the quantum field theories based on the concept of renormalization (which in a specific way includes both locality and unitary symmetry).

We shall suggest and apply at previously suggested experiments a model of the measurement or detection, i.e. collapse as spontaneous (non-dynamical) unitary symmetry, i.e. superposition and entanglement breaking (effective hiding) detailedly discussed in [24], [25] . It is in full agreement with all experimental data on the one hand, and, on the other hand, it consider that deterministic unitary symmetric quantum dynamics is sufficient for description of all exact quantum phenopmena, while collapse by detection (as spontaneous superposition breaking) describes effectively exact quantum phenomena. Here quantum mechanics represents a natural link between classical mechanics and quantum field theories.

Consider quantum mechanics as exact physical theory and its unitary symmetric dynamics as the exact dynamics.

Suppose that quantum dynamical state represents a wave packet. Under additional, well-known [2],[7],[29] approximation conditions mentioned wave packet can be globally approximately treated as the classical mechanical particle that satisfy approximate classical mechanical dynamical equation. (Namely, in Ehrenfest picture, average value of the quantum dynamical equation can be Taylor expanded [29] so that first term in expansion represents classical dynamical term for average coordinate value while other terms are proportional to increasing degrees of the coordinate uncertainty exponents. If coordinate average value is much larger than coordinate uncertainty, i.e. wave packet width, first term turns out in classical dynamical term for wave packet while other terms can be effectively neglected. Down limit of such wave packet approximation is characterized by Heisenberg uncertainty relations.) In this way (until all approximation conditions are satisfied) classical mechanics can be considered as approximate physical theory.

Consider exactly quantum mechanically two wave packets with practically the same (coordinate intervals) widths. It can be considered that mentioned two wave packets weakly interfere if distance between their centers is larger than one width.

Consider quantum dynamical state that represents (non-trivial) superposition of two weakly interfering wave packets. As it has been proved in [24],[25], such quantum superposition cannot be globally classical presented as the classical particle. (Namely, then mentioned Taylor expansion of Ehrenfest average value of the quantum dynamical equation becomes divergent.) Simply speaking, exactly existing superposition of two weakly interfering wave packets, approximately classically is globally non-stable and non-observable. But, of course, mentioned superposition is approximately classically locally stable and observable within any of two wave packets. Then, according to general formalism of the spontaneous symmetry breaking, here inherently probabilistically (with typical quantum mechanical probabilities) and spontaneously event of appearance of one or other wave packet becomes realized at classical level of analysis accuracy. However, at the quantum level of analysis accuracy superposition of weakly interfering wave packets stands conserved. For this reason, as it has been previously discussed, mentioned event appearance does not correspond to any real "choice" if this word "choice" should imply deterministic description of the event within dynamical superposition breaking. However, if by new detection only such quantum observables for which mentioned wave packets represents eigen states will be analyzed, classical level can be used as technically simple theoretical method, but it cannot be considered as any principal "choice". Then, according to determinism of the classical dynamics, new detector will detect the same wave packet which appears by the previous spontaneous superposition breaking. But if by new detection other (complementary) observables will be detected quantum superposition of weakly interfering wave packets must be used. Obviously all this represents an excellent model of the self-collapse at classical level of analysis accuracy.

Further, consider quantum dynamical state that represents (non-trivial) superposition of two wave packets. Suppose that initially mentioned wave packets are strongly interfering but that, according to deterministic quantum dynamical evolution, distance between centers of wave packets increases during time. Then in moment when wave packets centers become sufficiently distant, i.e. when wave packets become weakly interfering, conditions for self-collapse as spontaneous superposition breaking become satisfied and in this sense we can speak about collapse as a continuous phase transition.

Consider now measurement or detection procedure. Before quantum dynamical interaction between quantum system and detector, quantum system is described by superposition of the eigen states of measured observable, while detector is described by "zero" eigen state of the pointer observable. During deterministic quantum dynamical interaction between quantum system and detector entangled quantum state (bi-orthogonally expanded over quantum system measured observable eigen states and detector pointer observable eigen states) of the quantum super-system (that includes quantum system and detector) becomes realized. This entangled state (in absence of new interactions with additional physical systems) stands conserved during time. Suppose that within mentioned entangled state of super-system detector pointer observable eigen states represent wave packets. Suppose that mentioned wave packets are initially strongly interfering and that mentioned wave packets become during time weakly interfering. In moment when all mentioned wave packets become mutually weakly interfering at the classically described detector self-collapse as spontaneous (non-dynamical) entanglement breaking (effective hiding) appears. In other word, here spontaneous superposition breaking at super-system appears in full analogy with spontaneous superposition breaking at simple system. Simultaneously, for reason of correlations between detector pointer observable eigen states and quantum system measured observable eigen states, in respect to self-collapsed detector quantum system becomes effectively uniquely described by corresponding statistical mixture of the eigen states of measured observable. It this sense at quantum system relative collapse effectively appears as seemingly "absolute". But within exact entangled state of super-system statistical mixture by relative collapse of the quantum system is one of many possible second kind mixtures of this quantum system. Relative collapse at quantum system is effective but not absolute quantum phenomena and it occurs only in respect to classically self-collapsed detector. Super-system that includes quantum system and detector is exactly described by entangled state. If by a new sub-systemic detection at quantum system only, only such quantum observables compatible with previously detected observable will be analyzed, quantum
system can be effectively exactly described by statistical mixture chacarteristic for previous relative collapse. In this sense detections of the first and new detector are identical. But if by new detection other (complementary) observables will be detected entangled state of super-system, i.e. a different second kind mixture of the quantum system must be used. Obviously all this represents an excellent model of the measurement or detection if full agreement with all known experimental data.

## 5. Discussion of the experiments

Previously discussed model we shall now apply for explanation of previously suggested experiments (Figure 3. and Figure 4.).

In mentioned experiments excited state of two-level system is approximated with product of the state of single photon localized within mentioned two-level system and ground state of two-level system. It needs an additional argumentation. Firstly we can repeat well-known facts that for time independent Hamiltonian observable of quantum system its eigen values and eigen states are time independent. It implies according to standard quantum mechanical dynamics that such quantum system (without interactions with other physical systems) and initially prepared in some excited quantum state must stand in the same excited quantum state in any later time moment. But as it is well-known in the real experimental situations quantum system turns out from excited in some less excited or ground state in relatively short time interval in common with spontaneous emission of single photon. Thus within standard quantum mechanical dynamics for two-level system it is supposed that excited state |E> and ground state |G> represent two state from the same Hilbert space. After spontaneous photon emission excited state |E> turns out in state |G> |P> where |P> represents state of the spontaneously emitted photon. But obviously |G> |P> and |G> do not represent states in the same Hilbert space, which means that exactly speaking |E> and |G> do not represent states in the same Hilbert space. Moreover, it implies that excited state before spontaneous photon emission represents state of interacting two-level system and electromagnetic field formally denoted by |TLS,P (t)> whose exact dynamical evolution needs extension of the standard quantum mechanical dynamics by quantum field theory dynamics. But such extension can be approximately neglected in most usual quantum mechanical considerations [7]. All this points out that suggested approximation of the two-level system excited state during its initial stability by product of the state of the photon localized within two-level system and two-level ground state is rough but not inconsistent.

Further we shall additionally consistently approximate state of the photon localized within a two-level system by a wave packet with coordinate corresponding to two-level system position (and width corresponding to linear dimension of the two-level system) and with zero momentum (since it has been supposed that two-level system motion can be approximately neglected).

Now we can consider exact entangled state of two photons (6) realized immediately after quantum dynamical interactions between photon and both two-level systems. In this entangled state quantum system represents free propagating photon with two states, reflected and transmitted. Simultaneously, in this entangled state detector represents localized photon in two states, two wave packets, first one is localized within TLS2 and second one is localized within TLS1. According to suppositions, especially since two two-level systems are sufficiently distant, wave packets corresponding to localized photon are extremely weakly interfering. All this admits that at classically described localized photon self-collapse as spontaneous entanglement breaking occurs. Then simultaneously, in respect to self-collapsed localized photon, at free propagating photon relative collapse occurs. In experiment presented at Figure 3. mentioned relative collapse at free propagating photon will be detected by new sub-systemic detection by detection photo plate. But in experiment presented at Figure 4. exact entangled state of two photons (6), or entangled state of the photon and two-two level system (5), quantum dynamically evolves in non-entangled state (10) of two two- level systems and single free propagating photon. This photon described by superposition (11) finally, after new quantum dynamical interaction with beam splitter photon evolves in state (12).

## 6. Conclusions

In conclusion the following can be repeated and pointed out. In this work we suggest an original, realistic and simple, form of the experiment of interference of single photon at beam splitter. Here quantum entanglement (Schrödinger cat effect) between photon and photon trajectories detector (which includes two same simple two-level systems) is studied. It admits a consistent model of the detection or measurement procedure. Concretely here (self-)collapse at detector is consistently treated as the spontaneous (non-dynamical) breaking (effective hiding) of the unitary symmetry (superposition or entanglement) of quantum mechanical dynamics. It can be considered as an especial case of the general formalism of the spontaneous symmetry breaking.